\documentclass[aps,prb,twocolumn,superscriptaddress,showpacs]{revtex4}
\usepackage{graphicx}
\usepackage{helvet}

\begin{document}

\title{Emergence of inhomogeneous moments from spin liquid in the triangular-lattice Mott insulator $\kappa$-(ET)$_2$Cu$_2$(CN)$_3$}

\author{Y. Shimizu}
\altaffiliation[present address:]{RIKEN,Wako,Saitama 351-0198,Japan.}
\affiliation{Department of Applied Physics, University of Tokyo, Bunkyo-ku, Tokyo, 113-8656, Japan.}
\affiliation{Division of Chemistry, Kyoto University, Sakyo-ku, Kyoto, 606-8502, Japan.}
\author{K. Miyagawa}
\affiliation{Department of Applied Physics, University of Tokyo, Bunkyo-ku, Tokyo, 113-8656, Japan.}
\affiliation{CREST, Japan Science and Technology Agency, Kawaguchi, Saitama 332-0012, Japan.}
\author{K. Kanoda}
\affiliation{Department of Applied Physics, University of Tokyo, Bunkyo-ku, Tokyo, 113-8656, Japan.}
\affiliation{CREST, Japan Science and Technology Agency, Kawaguchi, Saitama 332-0012, Japan.}
\author{M. Maesato}
\affiliation{Division of Chemistry, Kyoto University, Sakyo-ku, Kyoto, 606-8502, Japan.}
\author{G. Saito}
\affiliation{Division of Chemistry, Kyoto University, Sakyo-ku, Kyoto, 606-8502, Japan.}

\date{\today}

\begin{abstract}
	The static and dynamic local spin susceptibility of the organic Mott insulator $\kappa$-(ET)$_2$Cu$_2$(CN)$_3$, a model material of the spin- 1/2 triangular lattice, is studied by $^{13}$C NMR spectroscopy from room temperature down to 20 mK. 
	We observe an anomalous field-dependent spectral broadening with the continuous and bipolar shift distribution, appearing without the critical spin fluctuations.
	It is attributable to spatially nonuniform magnetizations induced in the spin liquid under magnetic fields. 
	The amplitude of the magnetization levels off below 1 K, while the low-lying spin fluctuations survive toward the ground state, as indicated by the temperature profile of the relaxation rates. 
\end{abstract}



\maketitle

	Intensive research for exotic and unconventional superconductivity in layered cuprates and organic conductors has fertilized the physics of quantum antiferromagnets that may be a key to elucidate the mechanism of the electron pairing in the neighboring superconducting state and to find novel quantum states. 
	Of great interest is the realization of quantum spin liquids in two dimensions without any symmetry breaking since the proposal on the spin-1/2 triangular-lattice antiferromagnet (TLA).\cite{Anderson} 
	The succeeding theoretical studies of TLA, however, prefer a long-range spiral order at least in the Heisenberg model with nearest-neighbor exchange interactions \cite{Huse}, while the quantum disordered states with broken symmetry in translation,\cite{Sachdev} chirality \cite{Baskaran} and rotation \cite{Chandra} have been proposed. 
	The model including the higher-order exchanges \cite{Misguich, Motrunich} or the Hubbard model \cite{Imada} has predicted a quantum disordered ground state. 
	In contrast to the significant theoretical progress, a few model materials of the spin-1/2 TLA have been discovered to exhibit quantum disordered states only recently, including the $^3$He monolayer,\cite{Siqueira} Cs$_2$CuCl$_4$ \cite{Coldea} and organic conductors.\cite{Shimizu, Tamura} 

	The organic Mott insulator $\kappa$-(ET)$_2$Cu$_2$(CN)$_3$ is a promising system of the triangular-lattice spin liquid,\cite{Shimizu} where ET denotes the organic molecule, bis(ethylenedithio)tetrathiafulvalene. 
	The quasi-2D crystal structure consists of conducting ET layer and insulating Cu$_2$(CN)$_3$ layer. 
	In the conduction layer, strongly dimerized ET molecules are arranged in a checkerboard pattern, and the transfer integrals between the dimers form a triangular lattice with a half-filled band. 
	In fact, the magnetic susceptibility behaves as the TLA Heisenberg model with an exchange interaction energy $J$ of 250 K.\cite{Shimizu, Zheng} 
	Nevertheless, the long-range magnetic order was not observed down to 32 mK ($\sim$ 10$^{-4}J$) in the $^1$H NMR measurements.\cite{Shimizu} 
	Moreover, the residual spin susceptibility and the power-law temperature dependence in the spin-lattice relaxation rate at low temperatures imply the low-energy spin excitations. 
	Under a relatively low pressure, this material exhibits the Mott transition and the superconductivity without passing through the magnetically ordered state.\cite{Komatsu, Kurosaki} 
	The positive slope of the Mott transition phase boundary in the pressure-temperature phase diagram shows the large residual spin entropy in the insulating state.\cite{Kurosaki, Motrunich} 
	These results suggest that the novel quantum liquid with spin entropy not less than in Fermi liquid is realized near the Mott transition due to the significant spin frustration and higher-order or multi-particle exchange,\cite{Motrunich} and attracts great theoretical interests.\cite{Lee, Singh} 
	However, the nature of the spin liquid remains to be revealed experimentally, since the previously performed $^1$H NMR measurements could not resolve the local spin susceptibility due to the small electron density at the $^1$H sites. 

	In this paper, we report the $^{13}$C NMR measurements on $\kappa$-(ET)$_2$Cu$_2$(CN)$_3$, which have revealed the unusual inhomogeneous staggered fields emerging at low temperatures under external fields without an indication of the ordinary long-range order. 
	The inhomogeneous field develops with an increasing external field but saturates to a small value of moment at low temperatures under a fixed field, while the slow spin fluctuations are persistent. 
	The results show the partial breaking of the spin liquid by the external field and/or a possible local disorder. 

	The $^{13}$C NMR experiments were performed for a single crystal where $^{13}$C was enriched for the doubly bonded carbon sites at the center of ET.
	The hyperfine coupling constant at these sites is several tens times larger than at the $^1$H sites. 
	The NMR spectra were obtained by Fourier transformation (FT) of spin echo signals refocused after the pulse sequence of $\pi$/2-$\tau$-$\pi$, where $\tau$ was the rf pulse interval. 
	The spin-lattice relaxation time $T_1$ was obtained from the recovery of the integrated spin-echo intensity $I$ after the saturation. 
	We evaluated the spin-echo decay rate 1/$T_2$ from the width of the FT spectrum of $I(2\tau)$ which oscillates due to the dipolar coupling between the neighboring $^{13}$C spins in the ET molecule. 
	The experiments below 2 K were performed by utilizing a top-loading type of dilution refrigerator, where the sample was soaked into a liquid $^3$He-$^4$He mixture to avoid the rf heating. 

	\begin{figure}
	\includegraphics[scale=0.41]{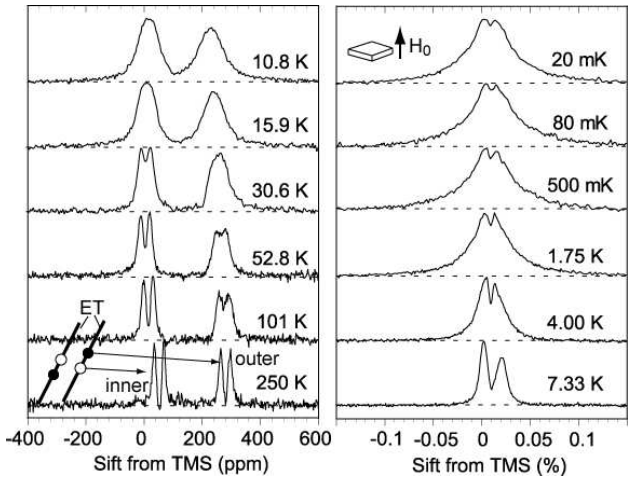}
	\caption{\label{Fig1} 
	$^{13}$C NMR spectra of $\kappa$-(ET)$_2$Cu$_2$(CN)$_3$ at $H_0$ = 8 T applied perpendicular to the conducting plane.
	}
	\end{figure}
	The crystal structure of $\kappa$-(ET)$_2$Cu$_2$(CN)$_3$ belongs to $P2_1/C$ in which all ET's are crystallographically equivalent. 
	When the magnetic field is applied perpendicular to the ET layer (parallel to the $a^*$ axis), all ET's are also magnetically equivalent.
	In fact, the observed four resonance lines, shown in Fig.1, at high temperatures are assigned to two neighboring and inequivalent $^{13}$C sites with the nuclear dipolar coupling in one molecule. 
	The two sets of doublets called the inner and outer $^{13}$C lines (see Fig.1) have hyperfine coupling constants $A_z$ of -0.07 and 0.21 T/$\mu_{\rm B}$dimer evaluated from the coefficient of the linearity between the Knight shift $K$ and the spin susceptibility $\chi_0$ in Ref. 11. 
	On decreasing temperature, the spectra are gradually broadened. 
	The similar behavior was also observed in the aligned polycrystal measurement of the singly $^{13}$C-enriched sample.\cite{Kawamoto} 
	The broadening is attributed to the hyperfine field of electron spins since the outer line with the larger $A_z$ is broader than the inner one. 
	Below 10 K, the inner and outer lines are merged due to a considerable decrease in the Knight shift. 
	In fact, it coincides with a decrease in $\chi_0$. 
	Nevertheless, the line broadening is still gradually developed and the tail edges reach $\pm$ 0.15\% in shift from the line center at 1.75 K, which corresponds to a maximum internal field of 12 mT. 
	At further low temperatures, however, the width and shape do not change significantly down to 20 mK. 
	The profile of the line broadening is quite unusual and distinct from that of the N\'eel order in $\kappa$-(ET)$_2$Cu[N(CN)$_2$]Cl which shows a clear line splitting by a huge internal field of 70 mT corresponding to the magnetic moment of 0.45-0.5$\mu_{\rm B}$/dimer.\cite{Miya2} 

	\begin{figure}
	\includegraphics[scale=0.4]{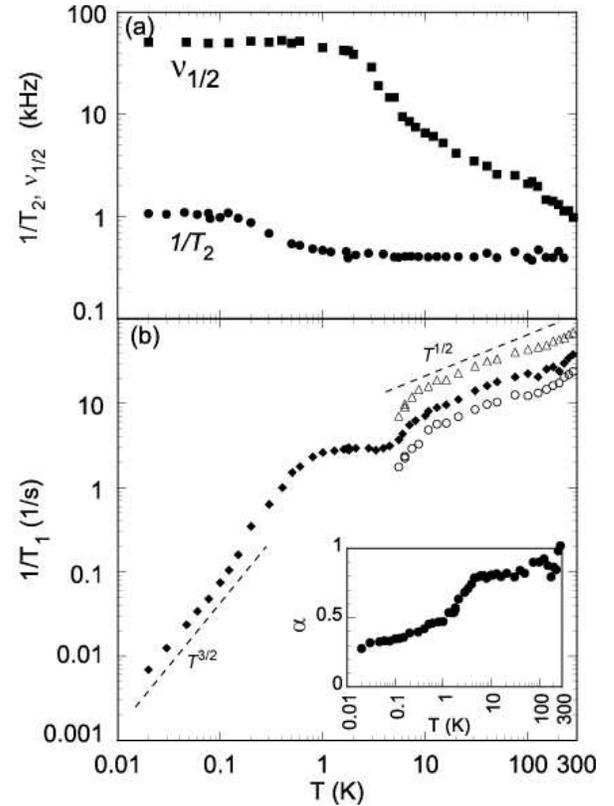}
	\caption{\label{Fig2} 
	(a) Temperature dependence of NMR linewidth $\nu_{1/2}$ and 1/$T_2$ measured at 8 T. 
	(b) 1/$T_1$ for the inner (triangle), outer (circle) and total $^{13}$C cites (diamond). 
	The inset shows the exponent $\alpha$ in the stretched exponential fit.
	}
	\end{figure}
	The spectral broadening generally indicates that the nuclear spins experience a static inhomogeneous field and/or a slowly fluctuating field from electron spins.
	The latter is probed by the spin-echo decay rate $1/T_2$, which characterizes field fluctuations causing irreversible dephasing of the nuclear spin polarization. 
	Figure \ref{Fig2} (a) shows $1/T_2$ and the spectral width $\nu_{1/2}$ defined by the full width at the half maximum of the outer $^{13}$C lines in the four-Lorentzian fitting of the spectra. 
	The $\nu_{1/2}$ grows far larger than $1/T_2$ particularly at low temperatures, indicating that the spectral broadening is ascribed primarily to the inhomogeneous broadening. 
	However, it is noted that $1/T_2$ does not show a peak to be observed in case of a magnetic transition but exhibits a small increase, followed by a saturation below 0.1 K. 
	It suggests that the slow spin fluctuations persist toward $T$ = 0. 

	The spin-lattice relaxation rate, 1/$T_1$, ($\propto T\sum_{\bf q}\chi^{\prime\prime}({\bf q})$) probes the low-energy spin excitation at the NMR frequency, where $\chi^{\prime\prime}({\bf q})$ is the imaginary part of the dynamic susceptibility at a wave vector ${\bf q}$. 
	We could evaluate 1/$T_1$ from the single-exponential recoveries for the inner and outer lines separately, or a sum of the two exponentials for the whole spectrum above 6 K. 
	As shown in Fig.\ref{Fig2}(b), 1/$T_1$ for the two $^{13}$C sites monotonously decreases with temperature, roughly following $T^{1/2}$ down to 10 K. 
	Since the uniform component $\chi({\bf q} = 0)$ only weakly depends on temperature above 10 K,\cite{Shimizu} the temperature dependence of $1/T_1T$ ($\sim$$T^{-1/2}$) reflects the development of ${\bf q} \neq 0$ components in $\chi^{\prime\prime}({\bf q})$. 
	However, the growth of $\chi^{\prime\prime}({\bf q})$ is unusually smooth compared with the square-lattice antiferromagnet La$_2$CuO$_4$ \cite{Imai} and the 1D antiferromagnet Sr$_2$CuO$_3$ \cite{Takigawa} in which $1/T_1T$ develops following $1/T$ on cooling as expected in the quantum critical regime.\cite{Chakravarty} 
	The distinct feature in the present case is considered as the consequence of the spin frustration that suppresses the growth of the long-range spin correlation at a finite ${\bf q}$.

	Below 10 K, the decrease in $1/T_1$ gets steep, coinciding with a rapid decrease in $\chi_0$.\cite{Shimizu} 
	This is a possible signature of the further growing of spin correlations toward the quantum disordered ground state or a precursor of the valence-bond solid formation. 
	Just after that, the nuclear magnetization recovery becomes nonexponential below 5 K but is well described by the stretched exponential, 1 - $M(t)/M(\infty)$ = exp(- ($t/T_1)^\alpha$), that would be applied to impurity-doped spin systems. 
	The exponent $\alpha$ would be unity for the uniform spin system but decrease with increasing inhomogeneity in $T_1$. 
	As seen in the inset of Fig.2(b), $\alpha$ deviates from unity even above 6 K. 
	This is because the magnetization $M(t)$ is evaluated as the overall integral of the outer and inner lines with the different $T_1$ values. 
	What is remarkable is a further decrease from 4 K, manifesting the emergence of inhomogeneity in this temperature region, where the spectral width is enhanced. 
	At the same time, 1/$T_1$ becomes nearly constant and then decreases again steeply well below 1 K, followed by a power decay ($\sim$ $T^{3/2}$) at the ultra-low temperatures. 
	Noticeably, the small $\alpha$ despite the long $T_1$ below 0.1 K represents that the inhomogeneity spreads much larger than molecular scale, otherwise $\alpha$ would approach again unity since the $T_2$ process averages out the different $T_1$'s as seen in the charge-ordered state.\cite{Miyagawa2} 

	\begin{figure}
	\includegraphics[scale=0.4]{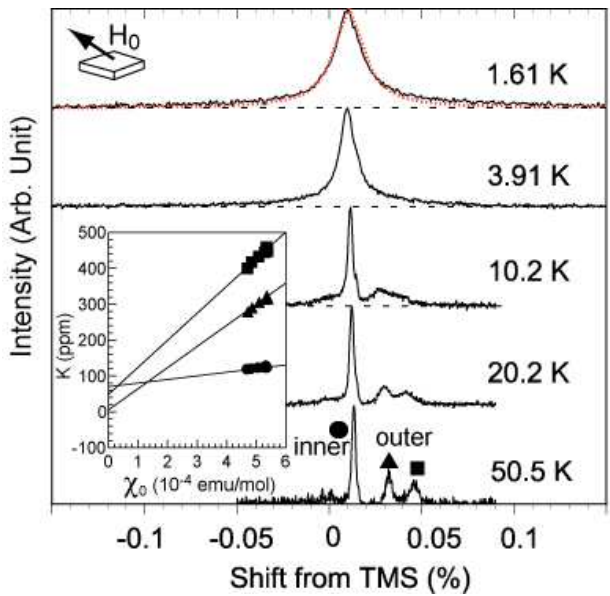}
	\caption{\label{Fig3} 
	The $^{13}$C NMR spectra of $\kappa$-(ET)$_2$Cu$_2$(CN)$_3$ under the external field of 8T applied at the magic angle. The red line is the model simulation of the spectrum. As for the model, see text.
	 }
	\end{figure}
	To explore the nature of the inhomogeneity, the external field was applied in a direction that makes all $A_z$'s positive and diminishes the nuclear dipole splitting. 
	Figure 3 displays the spectral profile and the $K-\chi_0$ plots of the three lines coming from the four inequivalent $^{13}$C sites in two ET's with different configurations against the external field; the two inner $^{13}$C lines are overlapped on the left line with the doubled intensity. 
	The $K-\chi_0$ plots show that all the three sites have positive $A_z$ and the Knight shift origins ($\chi_0$ = 0) are around 50($\pm50$)ppm. 
	These lines are broadened and merged into a single line on cooling due to the $\chi_0$ decrease. 
	Remarkably, the tails of the line are nearly symmetrical to the line center (at $\sim$80ppm) below 3.9 K. 
	The bipolar broadening points to the emergence of both the positive and negative spin polarization, namely, a staggered inhomogeneous field, instead of the paramagnetic inhomogeneity giving only a positive shift distribution. 
	The value of the magnetization is much smaller (less than 0.1$\mu_{\rm B}$) than the local moment in long-range order. 
	The Lorentzian-like spectral shape with long tails suggests that the larger magnetization is distributed in a far smaller volume fraction and the majority of electron spins carry almost uniform and small susceptibility. 
	The peak position of the spectra seems to have no resolved Knight shift below 4 K  but the shape is slightly asymmetric, implying that a small but finite spin susceptibility ($\sim2\times10^{-4}$ emu/mol) is remained at least at 1.6 K. 

	\begin{figure}
	\includegraphics[scale=0.35]{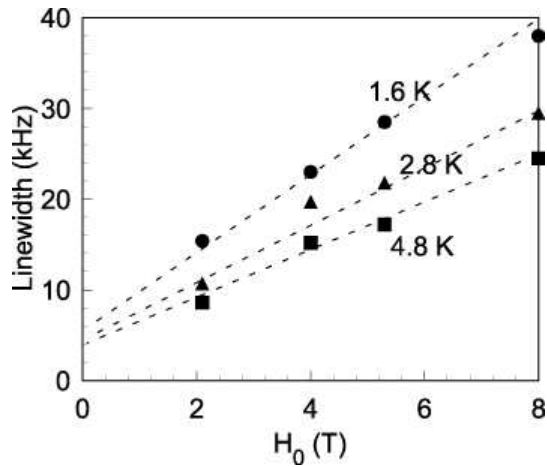}
	\caption{\label{Fig4} 
	External field dependence of the linewidth at various temperatures ($H_0$ $\parallel$ $a^*$).
	}
	\end{figure}
	The magnetic-field dependence of the linewidth is displayed in Fig.4. 
	The $\nu_{1/2}$ increases linearly with $H_0$ up to 8 T. 
	The extrapolation of the $\nu_{1/2}$ value to $H_0$ = 0 tends to about 5 kHz, which is nearly equal to the nuclear dipole field and indicates the vanishing internal field. 
	This agrees with the $\mu$SR measurements at $H_0$ = 0.\cite{Ohira} 
	These results show that the inhomogeneous magnetization appears under the external magnetic field, in contrast to the field-independent moment in the typical magnets. 

	The spatially-inhomogeneous magnetization at low temperature is thus quite anomalous in the sense that (i)the moments appear without a critical slowing down of the spin fluctuations, (ii)the magnetization grows linearly with the external field, (iii)the value of the moments saturates to a much smaller value than that of the classical value at low temperatures, and (iv)the slow spin fluctuations persist toward $T = 0$. 
	A conceivable origin of the inhomogeneous field is a spin glass state sometimes encountered in frustrated spin systems such as SrCr$_8$Ga$_4$O$_{19}$.\cite{Uemura} 
	In $\kappa$-(ET)$_2$Cu$_2$(CN)$_3$, however, the hysteresis of the magnetization against magnetic field and temperature, commonly seen in spin glass, was not observed down to 1.9 K above 0.3 T.\cite{Shimizu} 

	An alternative and likely origin that gives the above features is the magnetization nucleated around locally symmetry-broken sites, such as impurites or grain boundaries, as seen in 1D quantum antiferromagnets.\cite{Takigawa2} 
	In this case, the spatial distribution and the temperature dependence of the induced moment reflect the correlation function of the underlying spin liquid. 
	For an attempt to characterize the spatial variation of the moments, we made a simple model simulation for the NMR line shape $f(\omega)$ by a sum of the signals $g(\omega)$ with the internal fields from the staggered moments $S_i^z$ on the $i$-th concentric circles: $f(\omega)=\Sigma_i2\pi i g(\omega-(\omega_0 + A_zg\mu_{\rm B}S_i^z))$. 
	Here we assumed $A_z\sim$0.2 ${\rm T}/\mu_{\rm B}$, the concentration of the moment core of 0.01\%, the Lorentian $g(\omega)$ with the linewidth of 90ppm and an exponential decay $S_i^z=(-1)^iS_0$exp$(-i/\xi)$ with the correlation length $\xi$. 
	Then we obtained the successful reproduction of the line shape by adoping $S_0$ = 0.05$\mu_{\rm B}$/electron and $\xi$ = 10 lattices, as shown in Fig. 3. 
	The invariant line shape down to 20 mK suggest that the finite spin corelation is maintained in the underlying spin liquid, which will be compatible with the fully gapped state such as the chiral and dimer orders, or the nodally gapped one like the $d$-wave resonating valence bond (RVB). 
	The low-temperature saturation of $1/T_2$ is consistent with the finite spin correlation in the $T=0$ limit. 
	It is noted, however, that $1/T_2$ reflects $\chi({\bf q})$ perturbed by the symmetry-breaking sites, while the spatial moment distribution reflects $\chi({\bf q})$ before the perturbation. 
	This explains the different temperature dependences of the spectral width and 1/$T_2$. 
	It is not ruled out that the fully gapless nature underlies in the spin liquid as suggested theoretically \cite{Motrunich, Lee}, because the finite correlation of the moments can be deduced by the spin frustration and the screening effect analogous to the Friedel (or Ruderman-Kittel-Yoshida) oscillation in metals, even when the RVB spreads in a long range. 

	A recent study of the Heisenberg model with the ring-exchange interaction on the triangular lattice predicted that the inhomogeneous staggered magnetization appears even in a pure spin liquid with spinon Fermi surface when the external field couples to the spin chirality through the orbital motion.\cite{Motrunich2} 
	The detailed angular dependent experiments aginast the magnetic field are needed to confirm this proposal. 

	In conclusion, $^{13}$C NMR detected the anomalous field-induced inhomogeneous moments emerging on the triangular-lattice antiferromagnet $\kappa$-(ET)$_2$Cu$_2$(CN)$_3$ below 4 K. 
	The spin correlation of the inhomogeneous moments was likely saturated at low temperatures, while the slow spin fluctuations persisted toward $T=0$. 

	We thank A. Kawamoto for the synthesis of $^{13}$C substituted ET, and M. Tamura, M. Imada, N. Nagaosa, H. Fukuyama, R. McKenzie, and J. Zaanen for stimulating discussions. 
	This work was partially supported by MEXT KAKENHI on Priority Area of Molecular Conductors (No.15073204), JSPS KAKENHI (No.15104006) and COE project. 
	One of the authors (K.M.) is indebted to the Sumitomo Foundation for financial support.

\end{document}